# Investigation of activation cross sections of proton induced reactions on rhodium up to 70 MeV for practical applications

F. Tárkányi[1], F. Ditrói[1*], S. Takács[1], A. Hermanne[2], M. Baba[3], H. Yuki[3], A.V. Ignatyuk[4]

[1] *Institute for Nuclear Research, Hungarian Academy of Sciences (ATOMKI), Debrecen, Hungary*

[2] *Cyclotron Laboratory, Vrije Universiteit Brussel (VUB), Brussels, Belgium*

[3] *Cyclotron Radioisotope Center (CYRIC), Tohoku University, Sendai, Japan*

[4] *Institute of Physics and Power Engineering (IPPE), Obninsk, Russia*

**Abstract**

Excitation functions were measured for the production of the $^{101,100}$Pd, $^{102m,102g,101m,101g,100,99m,99g}$Rh and $^{97}$Ru radionuclides by bombardment of $^{103}$Rh targets with proton beams up to 70 MeV, some of them for the first time. The new results are compared with the earlier experimental data and with the theoretical nuclear model code calculations from ALICE-IPPE, EMPIRE and TALYS 1.6. Thick target yields were deduced and possible application of the new data for production of medically relevant $^{101m,101g}$Rh and $^{97}$Ru are discussed.

[*] Corresponding author: ditroi@atomki.hu

## 1. Introduction

In the frame of systematic study of activation cross sections of light charged particle induced nuclear reactions for different applications [1, 2] and for development of the reaction models, we have investigated the activation cross sections of proton induced nuclear reactions on rhodium (monoisotopic $^{103}$Rh) targets. In earlier studies we have already investigated the proton and deuteron induced activation cross sections of longer-lived products up to 40 and 50 MeV [3-5]. In one of our recent publications an overview of our and others' studies for production of clinically used and commercially available $^{103}$Pd was given [6] and recommendations for production are summarized in an IAEA technical report [7]. In this work we have extended the energy range of proton induced data for other radio-products up to 70 MeV. Only for a few of these reactions experimental cross section data are known (Scholz 1977 [8], Lagunas-Solar et al. [9-12], Hermanne et al. [3] and Sudar et al. [13]). Yield data at low energies were reported by Dmitriev et al. [14].

Rhodium and rhodium alloys are used in fusion, fission technology and nuclear transmutation, where secondary high energy protons are induced. Reliable high energy experimental data are important for these applications.

Nowadays medium energy accelerators were introduced in medical isotope production (cyclotrons and linear accelerators). Some radioisotopes can be produced only in this way and some others due to the smaller stopping power of high energy protons can be produced more effectively, compared to low energy accelerators. In present work the production of $^{97}$Ru, $^{101m}$Rh and $^{101g}$Rh are the related isotopes.

## 2. Experimental

The excitation functions for the $^{103}$Rh(p,x) reactions were measured at the cyclotron of Tohoku University (CYRIC, Sendai, Japan) using the stacked foil technique. The experimental method was similar to the techniques used in our numerous earlier investigations of charged particle induced nuclear reactions for different applications. One stack was irradiated using 70 MeV incident proton energy.

In both experiments natural, high purity Rh foils (Goodfellow >99.98%, thickness 12.3 µm) were assembled together with target foils of other elements (used in separate investigations) and with monitor and degrader foils.

The target stack was composed of a 19 times repeated sequence of Zr (103 µm), Rh (12.3 µm), Al (520 µm, degrader and monitor), Mn (10 µm), Al (520 µm, degrader and monitor) and Ag (52 µm) foils.

The Al monitors foils were used for determination of the beam intensity and energy by re-measuring the excitation function for the $^{27}$Al(p,x)$^{22,24}$Na reactions at CYRIC over the entire covered energy range and comparing with recommended values (details in the next section). The target stacks were irradiated in a Faraday-cup like target holder at 24 nA for 30 min.

The gamma activity of the majority of the produced radionuclides was measured with standard high purity Ge detectors coupled to acquisition/analysis software. No chemical separation was done and measurements were repeated several times up to months after EOB. Due to the high induced activity in the large number of simultaneously irradiated targets and resulting high dose for personnel the first measurements started about 1 day after EOB. The limited detector capacity has also a consequence that not all foils could be measured in an optimal way to get reliable information about radionuclides with half-lives shorter than 5 h. The second series of measurements of Rh foils was started 10 days after EOB.

## 3.  *Data processing*

Multiple independent γ-lines are available, allowing an internal consistency check of the calculated activities, for most of the radionuclides assessed. The decay modes and spectrometric characteristics (γ-lines energy and abundance, half-life) are summarized in Table 1 and were taken from the NUDAT2 data base [15].

The cross sections were calculated from the well-known activation formula. For some of the assessed radionuclides cumulative processes (decay of metastable states or parent nuclides) contribute to the production process.

The particle flux was initially derived from the total charge on target using a digital integrator. The incident beam energy was determined from the accelerator settings and the mean energy in each foil was calculated by the polynomial approximation for stopping of Andersen and Ziegler [16].

The initial beam energy and intensity were adjusted by comparing the excitation functions of the $^{27}$Al(p,x)$^{22,24}$Na reactions, re-measured over the whole energy range studied, with the recommended values in the updated version of IAEA-TECDOC 1211 [17] (Fig. 1). The uncertainty on the incident energy on the first foil of the stack was ±0.2 MeV. Taking into account the cumulative effects of possible variation on incident energy and thickness of the different targets, the uncertainty on the median energy in the last foil of the stack was about ± 1.2 MeV. The uncertainty on each cross-section was estimated in the standard way [18] by taking the square root of the sum in quadrature of all individual linear contributions,. The following individual uncertainties are included: determination of the peak areas including statistical errors (0.1-40 %), the number of target nuclei including non-uniformity (5 %), the detector efficiency (5 %) and incident particle intensity (7 %). The total uncertainty of the cross-section values was evaluated to vary from 8 to14 %, except for the cases where the statistical errors (low count rates) where high.

Table 1 Decay data of the investigated reaction products ( [15, 19]), the gamma-lines used for the final results are underlined.

| Nuclide Decay path | Half-life | $E_\gamma$ (keV) | $I_\gamma$ (%) | Contributing reaction | Q-value (keV) GS→GS |
|---|---|---|---|---|---|
| $^{101}$Pd<br>ε: 100 % | 8.47 h | 269.67<br><u>296.29</u><br>565.98<br>590.44 | 6.43<br>19<br>3.44<br>12.06 | $^{103}$Rh(p,3n) | -19522.39 |
| $^{100}$Pd<br>ε: 100 % | 3.63 d | 74.78<br>84.00<br><u>126.15</u><br>158.87 | 48<br>52<br>7.8<br>1.66 | $^{103}$Rh(p,4n) | -27797.1 |
| $^{102m}$Rh<br>IT: 0.233%<br>ε: 99.767%<br>140.7 keV | 3.742 y | 631.29<br><u>697.49</u><br>766.84<br>1046.59<br>1112.84 | 56.0<br>44.0<br>34.0<br>34.0<br>19.0 | $^{103}$Rh(p,pn) | -9318.94 |
| $^{102g}$Rh<br>β-: 22 5 %<br>ε: 78 5 % | 207.3 d | <u>468.58</u><br>739.5<br>1158.10 | 2.9<br>0.53<br>0.58 | $^{103}$Rh(p,pn) | -9318.94 |
| $^{101m}$Rh | 4.34 d | 306.86 | 81 | $^{103}$Rh(p,p2n) | -16759.86 |
| IT: 7.20 %<br>ε: 94.8 %<br>157.41 keV | | 545.12 | 4.3 | $^{101}$Pd decay | |
| $^{101g}$Rh<br>ε: 100 % | 3.3 y | 127.23<br>198.01<br>325.23 | 68<br>73<br>11.8 | $^{103}$Rh(p,p2n)<br>$^{101m}$Rh decay | -16759.86 |
| $^{100}$Rh<br>ε: 100 % | 20.8 h | 446.15<br>539.51<br>822.65<br>1107.22<br>1553.35 | 11.98<br>80.6<br>21.09<br>13.57<br>20.67 | $^{103}$Rh(p,p3n)<br>$^{100}$Pd decay | -26653.8 |
| $^{99m}$Rh<br>β$^+$: 7.3 %<br>ε: 92.7 %<br>64.3 keV | 4.7 h | 340.8<br>617.8<br>1261.2 | 69<br>11.8<br>10.9 | $^{103}$Rh(p,p4n)<br>$^{99}$Pd decay | -34735.34 |
| $^{99g}$Rh<br>β+: 100 % | 16.1 d | 89.76<br>353.05<br>528.24 | 33.4<br>34.6<br>38.0 | $^{103}$Rh(p,p4n) | -34735.34 |
| $^{97}$Ru<br>ε: 100 % | 2.83 d | 215.70<br>324.49 | 85.62<br>10.79 | $^{103}$Rh(p,2p5n)<br>$^{97}$Rh decay | -49554.32 |

Increase the Q-values if compound particles are emitted by: np-d, +2.2 MeV; 2np-t, +8.48 MeV; n2p-$^3$He, +7.72 MeV; 2n2p-α, +28.30 MeV.

Decrease Q-values for isomeric states with level energy of the isomer

## *4. Theoretical calculations*

For theoretical estimation the ALICE-IPPE [20] and EMPIRE [21] nuclear reaction model codes were used. Results of ALICE-IPPE for excited states were obtained by applying the isomeric ratios derived from the EMPIRE code to the total cross-sections calculated by ALICE.

For an additional comparison with the experimental values the results of calculations with the latest version of the modified TALYS code ( Koning 2012 [22]) taken from the on-line TENDL-2014 [23] and TENDL-2015 [24] libraries are given.

## *5. Results and discussion*

### **5.1 Excitation functions**

The investigated cross-sections for radionuclides produced in the bombardment of $^{103}$Rh with protons are tabulated in Tables 2 and 3 respectively, and for comparison with the theory and with the earlier experimental results are shown in Figs 2-11 graphically. The Q-values of the contributing processes are collected in Table 1. We did not obtain reliable data for production of $^{103}$Pd, due to the missing low energy X-ray part in our gamma-spectra and to the low emission probability of the 357 keV γ-line. During the cooling time between EOB and measurement and in case of some longer-lived products the data are scattered due to the low counting statistics.

Table 2. Activation cross-sections for formation of $^{101,100}$Pd, $^{102m, 102g}$Rh reactions.

| E | ΔE | $^{101}$Pd | | $^{100}$Pd | | $^{102m}$Rh | | $^{102g}$Rh | |
|---|---|---|---|---|---|---|---|---|---|
| | | σ | Δσ | σ | Δσ | σ | Δσ | σ | Δσ |
| MeV | | mb | | | | | | | |
| 69.81 | 0.20 | 47.10 | 5.60 | 66.73 | 29.65 | 57.42 | 21.96 | | |
| 66.74 | 0.27 | 49.65 | 5.69 | 76.65 | 9.20 | 76.65 | 8.29 | | |
| 63.86 | 0.33 | 58.92 | 6.73 | 85.51 | 10.18 | 75.44 | 22.00 | | |
| 60.53 | 0.40 | 59.87 | 6.84 | 100.46 | 11.72 | 74.93 | 22.16 | | |
| 57.39 | 0.47 | 60.58 | 7.09 | 120.83 | 13.94 | 64.46 | 22.23 | | |
| 53.74 | 0.55 | 73.87 | 8.37 | 178.24 | 19.91 | 65.64 | 25.95 | 115.46 | 39.63 |
| 51.17 | 0.61 | 84.23 | 9.71 | 221.01 | 24.50 | 61.28 | 23.76 | | |
| 49.04 | 0.66 | 93.28 | 10.32 | 253.12 | 28.06 | | | | |
| 47.23 | 0.70 | 106.11 | 11.92 | 282.44 | 31.17 | | | | |
| 44.94 | 0.75 | 124.63 | 13.87 | 290.95 | 31.98 | 86.76 | 23.51 | | |
| 42.99 | 0.79 | 155.30 | 17.10 | 278.48 | 30.68 | 65.57 | 9.62 | 89.76 | 15.52 |
| 40.52 | 0.84 | 223.66 | 24.37 | 266.27 | 29.31 | 111.70 | 112.35 | | |
| 38.38 | 0.89 | 293.06 | 31.80 | 200.72 | 22.23 | 102.79 | 21.09 | | |
| 35.65 | 0.95 | 420.50 | 45.58 | 129.8 | 14.77 | 111.11 | 21.59 | | |
| 33.27 | 1.00 | 465.32 | 50.42 | 72.30 | 8.90 | 103.57 | 16.56 | | |
| 30.17 | 1.07 | 472.61 | 51.21 | 50.81 | 6.83 | 118.93 | 16.09 | 112.07 | 38.17 |
| 27.42 | 1.13 | 421.82 | 45.72 | 16.85 | 2.31 | 117.96 | 17.16 | 135.76 | 26.05 |
| 23.77 | 1.21 | 189.89 | 20.61 | | | 109.19 | 13.29 | | |
| 20.39 | 1.29 | 19.01 | 2.13 | | | 63.92 | 7.17 | 77.70 | 8.93 |

Table 3. Activation cross-sections for formation of $^{101m,101g,100,99m,99g}$Rh and $^{97}$Ru reactions.

| | | $^{101m}$Rh | | $^{101g}$Rh | | $^{100}$Rh | | $^{99m}$Rh | | $^{99g}$Rh | | $^{97}$Ru | |
|---|---|---|---|---|---|---|---|---|---|---|---|---|---|
| E | ΔE | σ | Δσ | σ | Δσ | σ | Δσ | σ | Δσ | σ | Δσ | σ | Δσ |
| MeV | | mb | | | | | | | | | | | |
| 69.81 | 0.20 | 149.62 | 18.53 | 92.37 | 10.52 | 157.06 | 1.30 | 164.55 | 18.27 | 50.92 | 5.53 | 21.19 | 2.33 |
| 66.74 | 0.27 | 153.61 | 19.10 | 53.82 | 8.16 | 152.76 | 1.19 | 193.23 | 21.13 | 56.73 | 6.97 | 22.13 | 2.43 |
| 63.86 | 0.33 | 161.03 | 19.78 | 59.09 | 10.06 | 161.64 | 1.14 | 204.68 | 22.33 | 49.93 | 6.94 | 25.06 | 2.74 |
| 60.53 | 0.40 | 165.26 | 20.63 | 61.55 | 13.23 | 184.86 | 1.08 | 209.92 | 22.92 | 47.66 | 5.21 | 32.80 | 3.58 |
| 57.39 | 0.47 | 177.02 | 21.74 | 62.52 | 12.94 | 203.73 | 1.10 | 182.74 | 20.27 | 39.75 | 4.37 | 40.70 | 4.43 |
| 53.74 | 0.55 | 193.26 | 23.73 | 78.38 | 15.73 | 261.93 | 0.89 | 141.09 | 15.62 | 29.23 | 3.26 | 57.99 | 6.30 |
| 51.17 | 0.61 | 216.91 | 27.10 | 85.27 | 18.08 | 293.78 | 0.91 | 78.16 | 9.65 | 16.10 | 1.90 | 61.95 | 6.73 |
| 49.04 | 0.66 | 218.88 | 26.90 | | | 277.99 | 0.83 | 40.44 | 4.71 | 1.26 | 0.25 | 59.45 | 6.47 |
| 47.23 | 0.70 | 248.69 | 29.74 | | | 291.54 | 0.90 | 18.69 | 4.29 | | | 53.50 | 5.82 |
| 44.94 | 0.75 | 267.31 | 31.49 | | | 249.78 | 0.94 | | | | | 42.00 | 4.58 |
| 42.99 | 0.79 | 356.04 | 40.36 | 95.02 | 11.23 | 209.61 | 1.01 | | | 1.35 | 0.23 | 30.08 | 3.30 |
| 40.52 | 0.84 | 372.90 | 41.58 | 152.08 | 16.68 | 139.22 | 1.18 | | | | | 15.22 | 1.71 |
| 38.38 | 0.89 | 502.05 | 55.44 | 156.58 | 19.33 | 62.75 | 1.48 | | | | | 5.12 | 0.65 |
| 35.65 | 0.95 | 660.24 | 72.29 | 108.19 | 20.09 | 17.71 | 2.44 | | | | | 0.80 | 0.35 |
| 33.27 | 1.00 | 653.96 | 71.57 | 89.62 | 19.21 | | | | | | | | |
| 30.17 | 1.07 | 626.74 | 68.62 | 91.70 | 16.87 | | | | | | | | |
| 27.42 | 1.13 | 534.98 | 58.68 | 37.43 | 13.23 | | | | | | | | |
| 23.77 | 1.21 | 221.17 | 24.71 | | | | | | | | | | |
| 20.39 | 1.29 | 200.63 | 27.89 | 4.74 | 0.95 | | | | | | | | |

### 5.1.1 The $^{103}$Rh(p,3n)$^{101}$Pd reaction

The reaction cross sections for formation of $^{101}$Pd ($T_{1/2}$ = 8.47 h) are shown in Fig. 2, together with the earlier experimental results and theoretical estimations. The new experimental data are in good agreement with our earlier results in magnitude. The Lagunas-Solar data [11] near the threshold are lower than our new data and the Sudar data [13] are somewhat higher. The Lagunas-Solar data show rather an energy shift and the Sudar data rather higher than ours below 33 MeV. A possible reason could be the incorrect energy scale and incorrect intensity at the stack end, respectively. Except for ALICE-IPPE, the other theoretical results agree in magnitude but show energy shifts.

### 5.1.2 The $^{103}$Rh(p,4n)$^{100}$Pd reaction

According to Fig. 3 there is good agreement with the earlier experimental data for the activation cross section data of $^{100}$Pd ($T_{1/2}$ = 3.63 d) [12, 13]. The results of the three codes show deficiencies in energy scale and maximal cross-section values.

### 5.1.3 The $^{103}$Rh(p,pn)$^{102m}$Rh and $^{103}$Rh(p,pn)$^{102g}$Rh reactions

The cross section values for both isomers separately are given in Fig. 4 and Fig. 5 in comparison with the theoretical results and with the results of Hermanne [3]. The very low isomeric transition ratio (IT: 0.233 %) and the long half-life of the $^{102m}$Rh ($T_{1/2}$ = 3.742 y) allow to separate the direct formation also of the ground state ($T_{1/2}$ = 207.3 d).

### 5.1.4 The $^{103}$Rh(p,p2n)$^{101m}$Rh and $^{103}$Rh(p,p2n)$^{101g}$Rh reactions

The radionuclide $^{101}$Rh has an excited state with a half-life of 4.34 d that decays for 7.20% by internal transition to the long-lived ($T_{1/2}$ = 3.3 y) ground state. The different independent γ-lines emitted during the decay of excited and ground-state allow to determine the activity of both isomers at different times after EOB. Due to the fact that $^{101m}$Rh is also formed by decay of the much shorter-lived $^{101}$Pd ($T_{1/2}$ = 8.47 h, 100%) we could only assess the activities of both $^{101}$Rh isomers in the measurement at EOB + 10 d. The cross-section for $^{101m}$Rh presented here is hence cumulative (Fig. 6) while the values for $^{101g}$Rh are also cumulative including complete decay of the shorter-lived metastable state (Fig. 7). Our measurements above 30 MeV for the metastable state are clearly higher and less peaked than the Lagunas-Solar data [11]. The TENDL (cumulative) predictions are rather well describing the experiment be it with a 7-8 MeV downshift in the position of the maximum. EMPIRE and especially ALICE-IPPE results are too high. The TENDL predictions for the direct production are about one third of the cumulative cross-section.

For the ground state our values form a straightforward continuation at higher energy of the Hermanne's data [3] and are a factor of five higher than results of Lagunas-Solar [11]. EMPIRE code gives the best description.

### 5.1.5 The $^{103}$Rh(p,p3n)$^{100}$Rh reaction

The presented cross-section of $^{100}$Rh ($T_{1/2}$ = 20.8 h) (Fig. 8) ground state includes the contribution from the not measured short-lived $^{100m}$Rh isomer ($T_{1/2}$ = 4.7 min)(m+), but the contribution of the decay from $^{100}$Pd ($T_{1/2}$ = 3.63 d) parent isotope was separated (Fig. 8). A reasonable agreement with Lagunas-Solar [11] is seen and all codes give acceptable predictions.

### 5.1.6 The $^{103}$Rh(p,p2n)$^{99m}$Rh and $^{103}$Rh(p,p2n)$^{99g}$Rh reactions

The radionuclide $^{99}$Rh has an excited state with a half-life of 4.7 h and a 16.1 d half-life ground state that are decaying independently. Both states are formed directly and through the decay of $^{99}$Pd ($T_{1/2}$ = 21. 4 min) (in 32.2% to the $^{99m}$Rh and in 67.78 % to the $^{99g}$Rh). The cumulative cross sections are shown in Figs. 9 and 10. No earlier experimental data have been found in the literature. For the metastable state the TENDL predictions are shifted to lower energies (there is also a shift between the both TENDL versions) and are 20% lower, while EMPIRE is shifted to higher energy. ALICE-IPPE is too high. For the ground state EMPIRE is rather well describing the experimental values, TENDLs are shifted too lower energies and ALICE-IPPE is predicting a double maximum cross-section.

5.1.7 The $^{103}$Rh(p,x)$^{97}$Ru reaction

Our measured cross sections for $^{97}$Ru ($T_{1/2}$ = 2.83 d) formation include the direct production by the $^{103}$Rh(p,$\alpha$3n) reaction with an energy threshold of 21.46 MeV and at higher energies by the $^{103}$Rh(p,2p5n) reaction (threshold = 50.04 MeV) combined with possible contribution of the $^{97}$Pd ($T_{1/2}$ = 3.10 min) – $^{97m,g}$Rh ($T_{1/2}$ = 30.7 min, 46.2 min) parent decay chain from about 53 MeV on ($^{103}$Rh(p,d5n)$^{97}$Rh reaction). The agreement with the earlier experimental data is good.

The two TENDL predictions underestimate the experimental values in the measured energy range and have a different shape above 60 MeV (Fig. 11). The theory also represents well the start on the contribution of the parents decay chain above 60 MeV. EMPIRE and ALICE-IPPE overestimate the direct production and have a very different behavior in the high energy part.

## 5.2 Integral yields

The integral yields for the $^{101,100}$Pd, $^{102m, 102g.101m,101g,100,99m,99g}$Rh and $^{97}$Ru reactions were calculated from the measured excitation functions. The results for physical yields [25] are shown in Fig. 12. Several authors [11, 26] published thick target yield measurement/calculation results for different energy regions, which are not directly comparable with our displayed results.

## 6. *Application of experimental data: production of $^{101m}$Rh, $^{101g}$Rh and $^{97}$Ru*

Apart from the established use of $^{103}$Pd for brachytherapy seeds at least three of the investigated reaction products have potential use in nuclear medicine: $^{97}$Ru and $^{101m,g}$Rh among others were recently mentioned for use in nuclear medicine in a review on the application of the radioisotopes of second transition series elements [27-29] (indirect production of the $^{101m}$Rh through the $^{101}$Pd parent and cumulative production of $^{97}$Ru via $^{103}$Rh(p,x) reaction)
The long-lived $^{101g}$Rh is also mentioned as X-ray source for crystallinity studies and elemental analysis by using XRD and XRF methods [30].

## 6.1 Production of $^{97}$Ru

Production of $^{97}$Ru is possible at reactors and accelerators. The $^{96}$Ru(n,γ) route results in no carrier free products. At accelerators many options were proposed and investigated: $^{99}$Tc(p,3n) [31, 32], $^{nat}$Mo($^3$He,xn) [33], $^{nat}$Mo(α,xn) [34-37], the here presented $^{103}$Rh(p,x) path [3, 9] and Y-targets with heavier ions [38].

The $^3$He and heavy ions are not of practical importance for medical radioisotopes production due to the low yield and the low availability of proper accelerators with high intensity beams.

The use of alpha particle beams is meaningful only when no alternative proton or deuteron induced routes exists and are rarely used because of the lower yield and availability of high intensity alpha particles. At present, high intensity proton induced reactions in the 70 - 40 MeV energy range are hence the most powerful for production of no carrier added $^{97}$Ru. High intensity proton beams in this energy range are already available both in new H$^-$ - cyclotrons and at linear accelerators. Deuteron induced reactions on Rh require even higher energy, which is practically not available for routine production. In our latest investigations up to 40 MeV no activity of $^{97}$Ru was detected [4] and in the 50 MeV irradiation [5] the production cross section reaches about 50 mb at 50 MeV (practical threshold is around 35 MeV)

## 6.2 Production routes of $^{101m}$Rh

*Non carrier added* $^{101m}$Rh can be produced directly via proton [14, 39-41] and deuteron [4, 5] induced reactions on ruthenium.

We investigated here an alternative route: indirect formation through the $^{101}$Pd parent 'generator'. The $^{101}$Pd is decaying with 99.75 % to the $^{101m}$Rh. Several low energy routes for production of $^{101}$Pd are documented, namely proton [2, 11, 26, 42] and deuteron [4, 5] induced reaction on $^{103}$Rh, proton [42, 43] and deuteron [44] induced reactions on palladium, alpha- and $^3$He-particle [45] induced reactions on ruthenium. The $^{103}$Rh(p,n) reaction has high production yields. However, to avoid contamination of the $^{100}$Rh the proton reaction could be used below 30 MeV i.e. below the threshold of the $^{103}$Rh(p,4n) reaction.

## 6.3 Production routes of $^{101g}$Rh

The longer-lived ground-state can be produced by identical reactions used for direct production of $^{101m}$Rh and as a decay product of the metastable state (IT 9.7 %). A large contamination with $^{101m}$Rh will always be present (higher cross-sections, shorter half-life) at EOB, but due to the long half-life of the ground state long cooling-time is allowed, two months cooling will reduce the initial level with a factor of 10000.

## 7. Summary

Activation cross sections were measured for production of the $^{101,100}$Pd, $^{102m,102g,101m,101g,100,99m,99g}$Rh and $^{97}$Ru radionuclides by proton induced reactions on $^{103}$Rh targets for the first time, for production of $^{102m,102g}$Rh above 30 MeV and of $^{99m,99g}$Rh in the whole investigated energy range. The new experimental data are in good agreement with the previous low energy data. The agreement is not so good for some reactions of earlier experimental data obtained at high energy irradiation. The TENDL has usually acceptable predictions for proton induced reactions, but in case of the recent reactions in some cases there are large disagreements on the magnitude, especially in case of isomers. The quality of the description in case of the EMPIRE in several cases is better than the TENDL predictions and the ALICE-IPPE code gives the worst results in most cases. The comparison shows the importance of the experimental data.

**Acknowledgement**

This work was performed in the frame of the collaboration of the ATOMKI and CYRIC institutes (Hungary-Japan). The authors acknowledge the support of the research project and of the respective institutions in providing the beam time and experimental facilities.

**Figures**

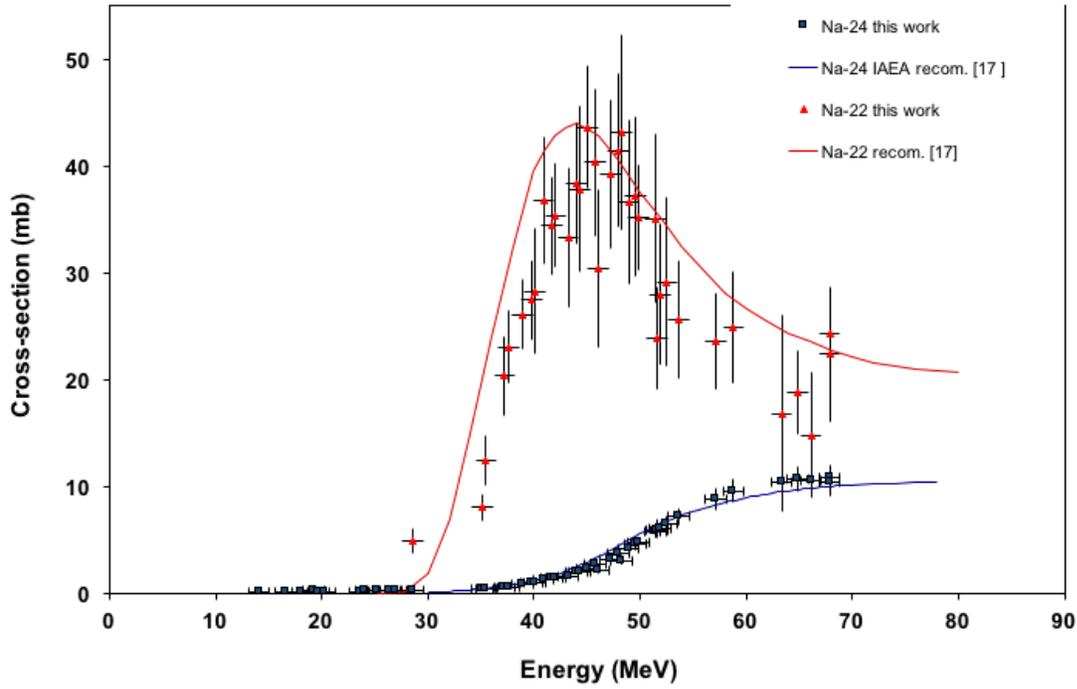

Fig.1 The re-measured cross sections of the used monitor reactions in comparison with the recommended data

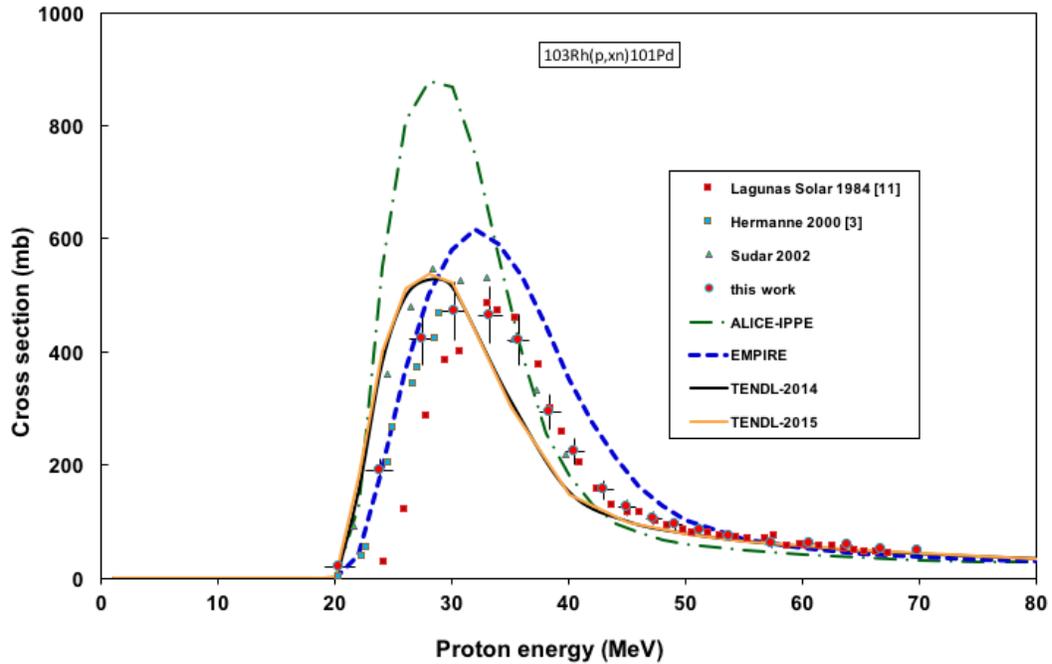

Fig.2. Excitation function of the $^{103}$Rh(p,x)$^{101}$Pd reaction compared with the theory and literature

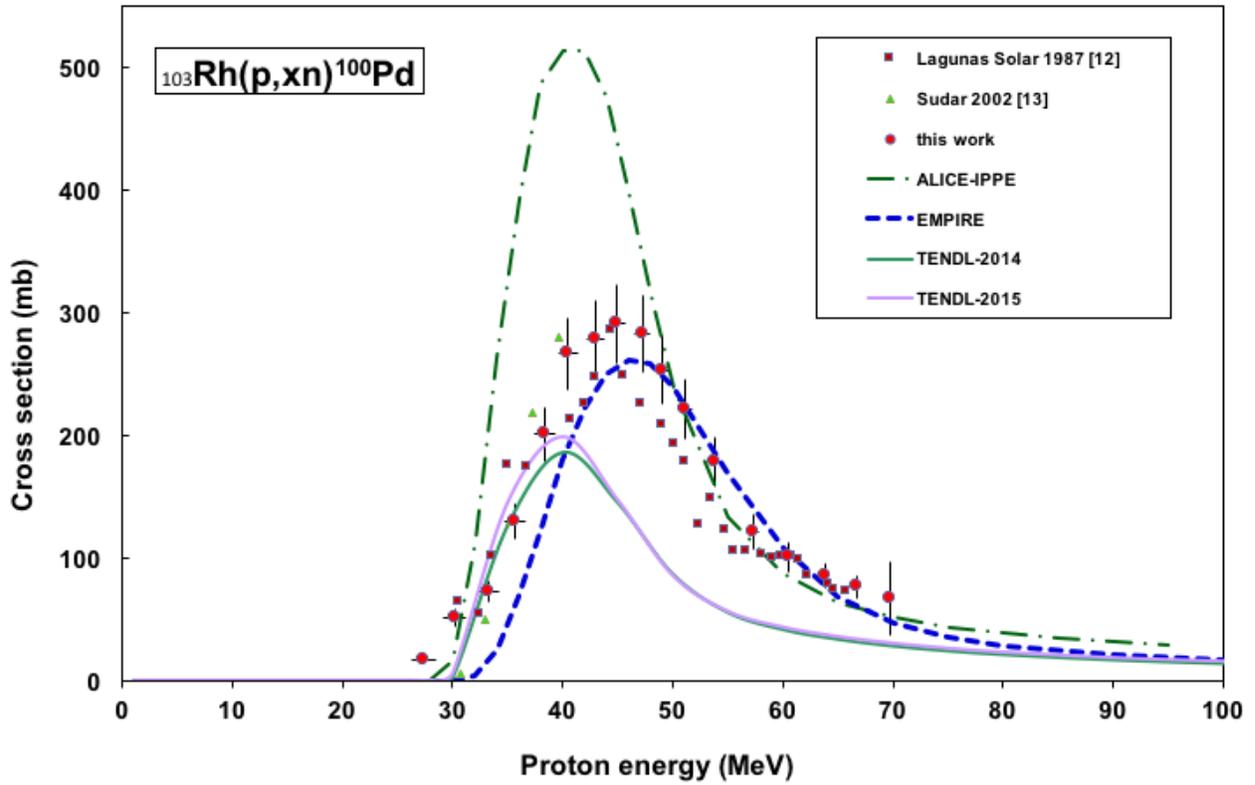

Fig.3. Excitation function of the $^{103}$Rh(p,x)$^{100}$Pd reaction compared with the theory and literature

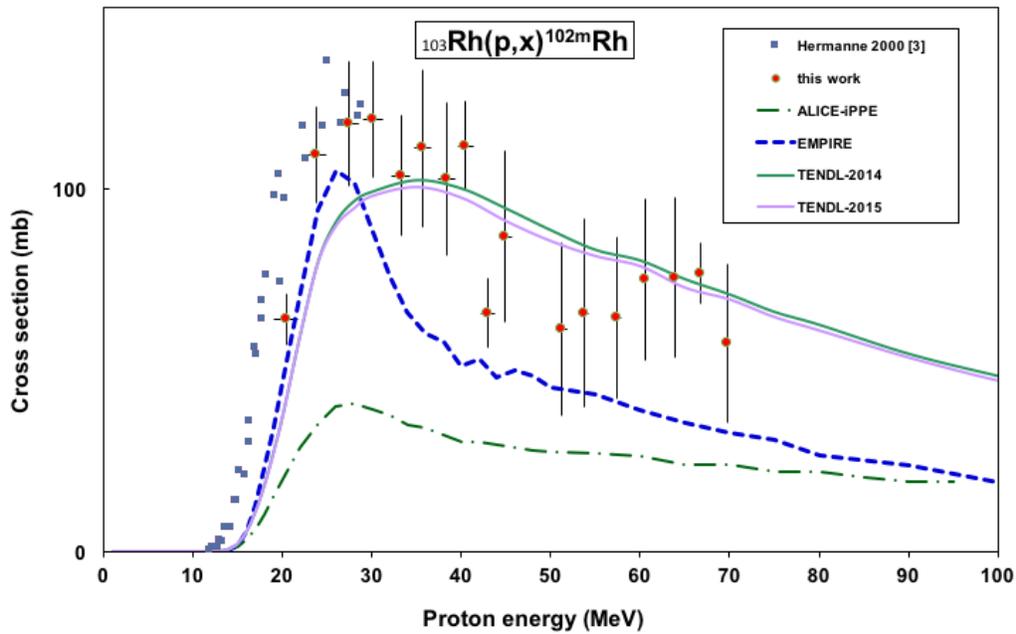

Fig.4. Excitation function of the $^{103}$Rh(p,pn)$^{102m}$Rh reaction compared with the theory and literature

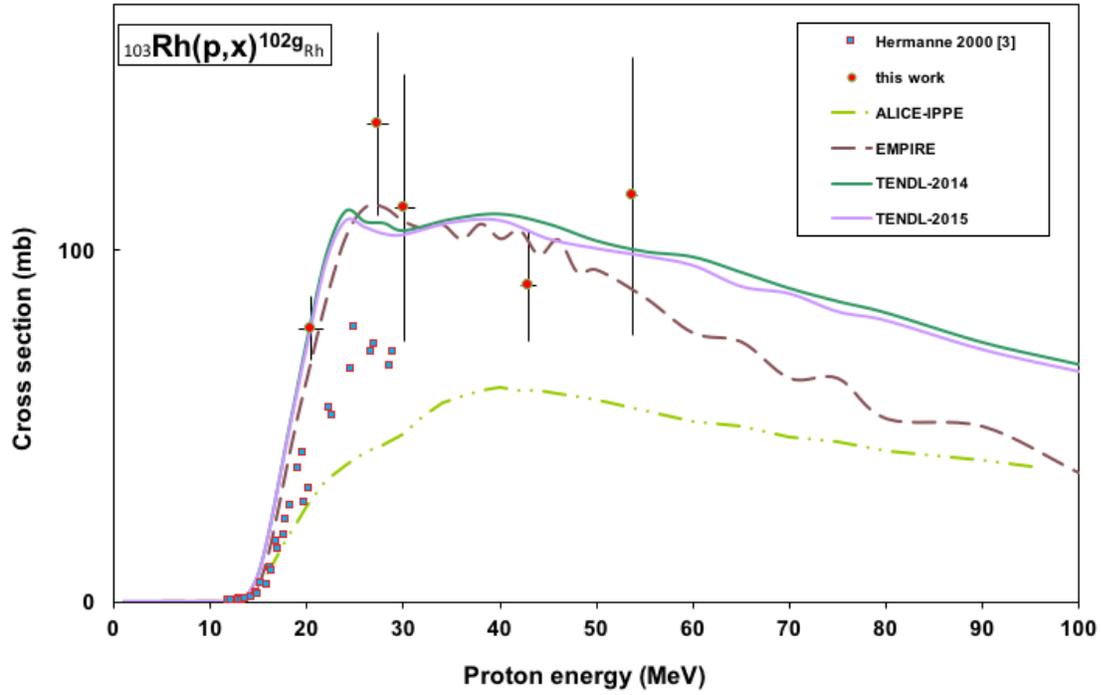

Fig.5. Excitation function of the $^{103}$Rh(p,pn)$^{102g}$Rh reaction compared with the theory and literature

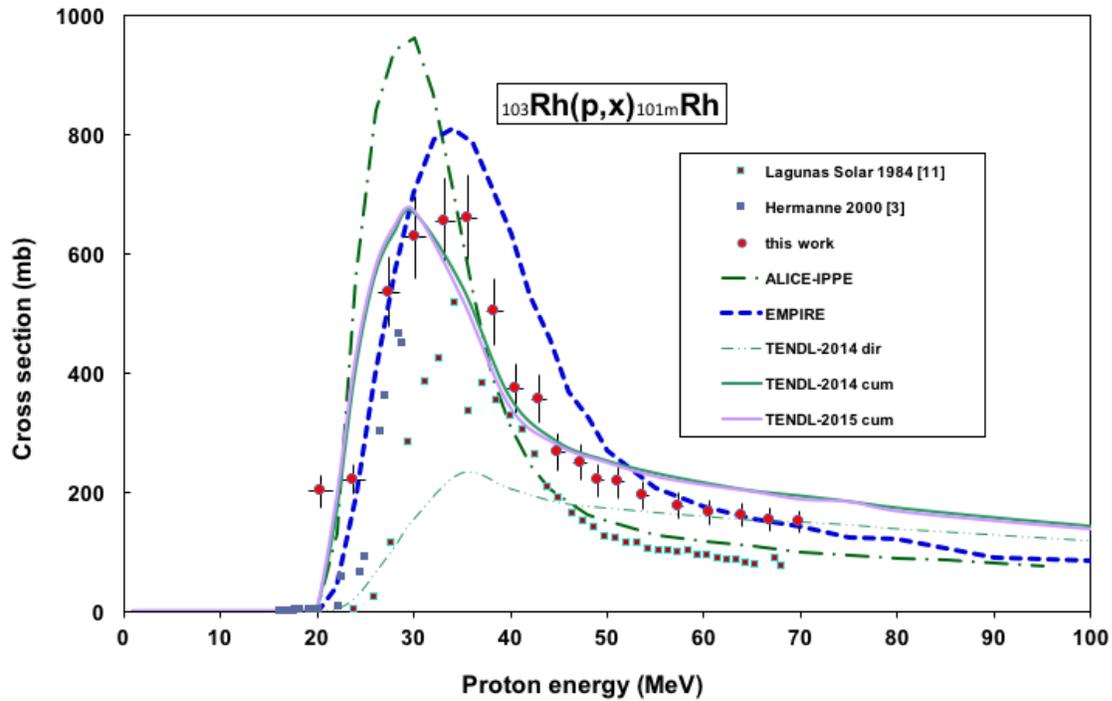

Fig.6. Excitation function of the $^{103}$Rh(p,pn)$^{101m}$Rh reaction compared with the theory and literature

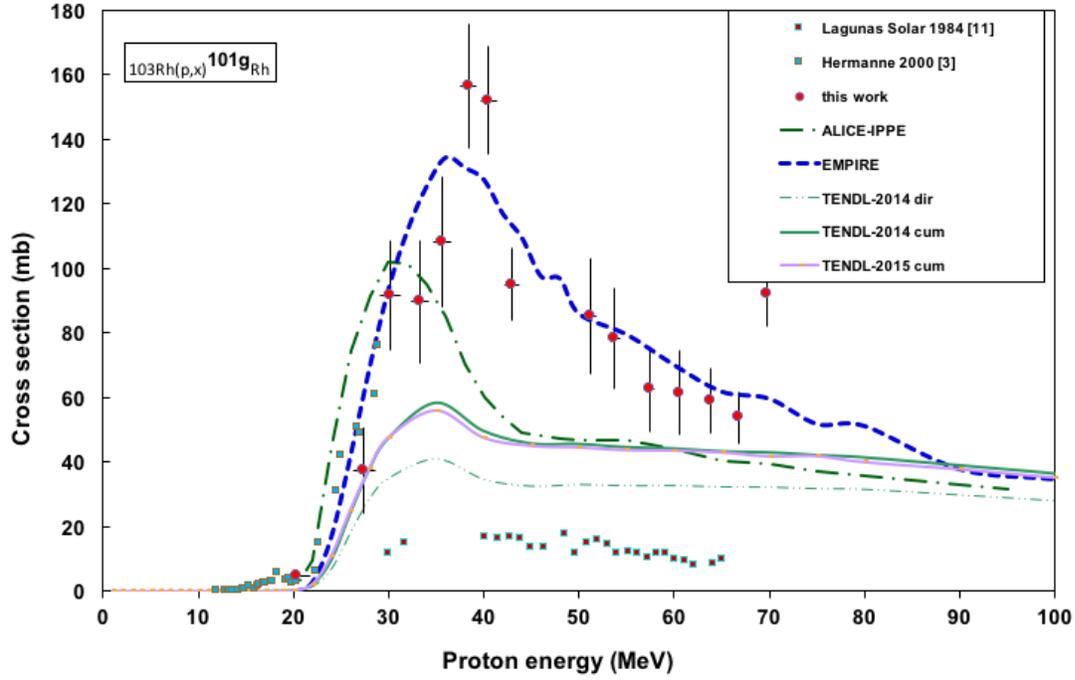

Fig.7. Excitation function of the $^{103}$Rh(p,pn )$^{101g}$Rh reaction compared with the theory and literature

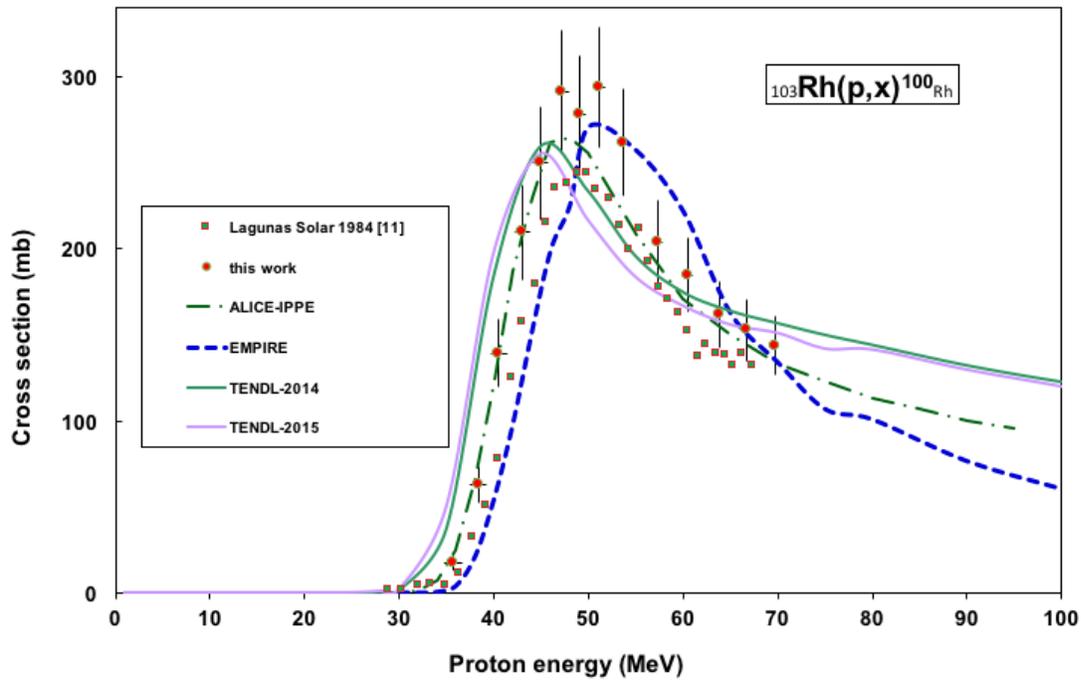

Fig.8. Excitation function of the $^{103}$Rh(p,pn )$^{100}$Rh reaction compared with the theory and literature

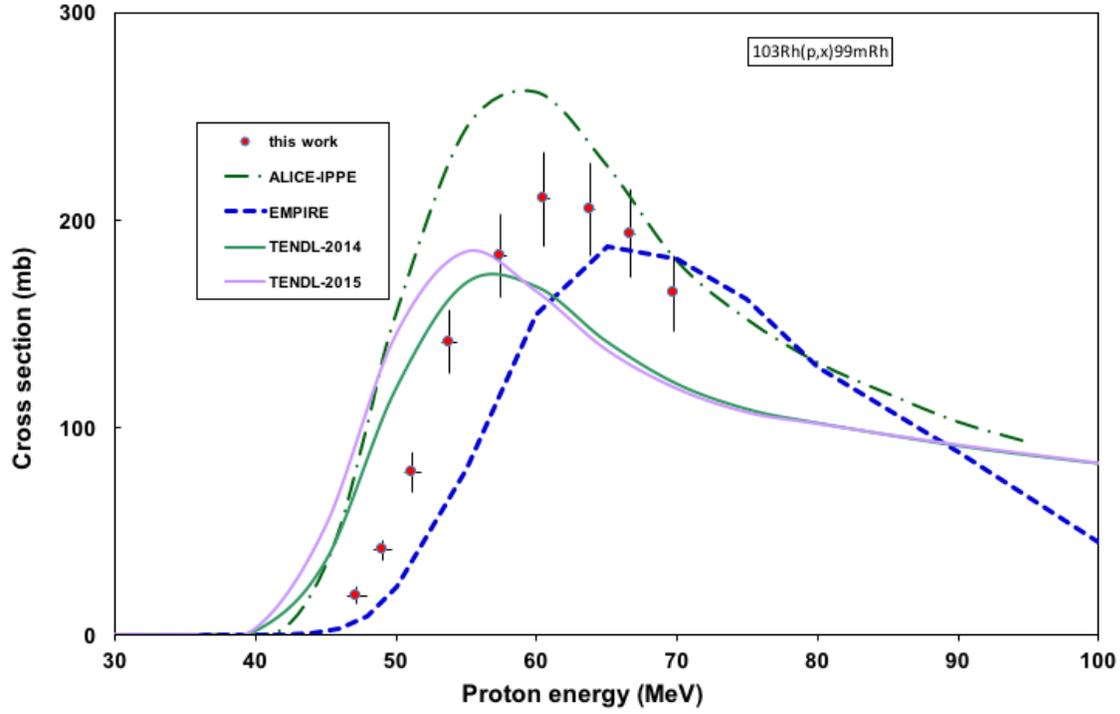

Fig.9. Excitation function of the $^{103}$Rh(p,pn)$^{99m}$Rh reaction compared with the theory and literature

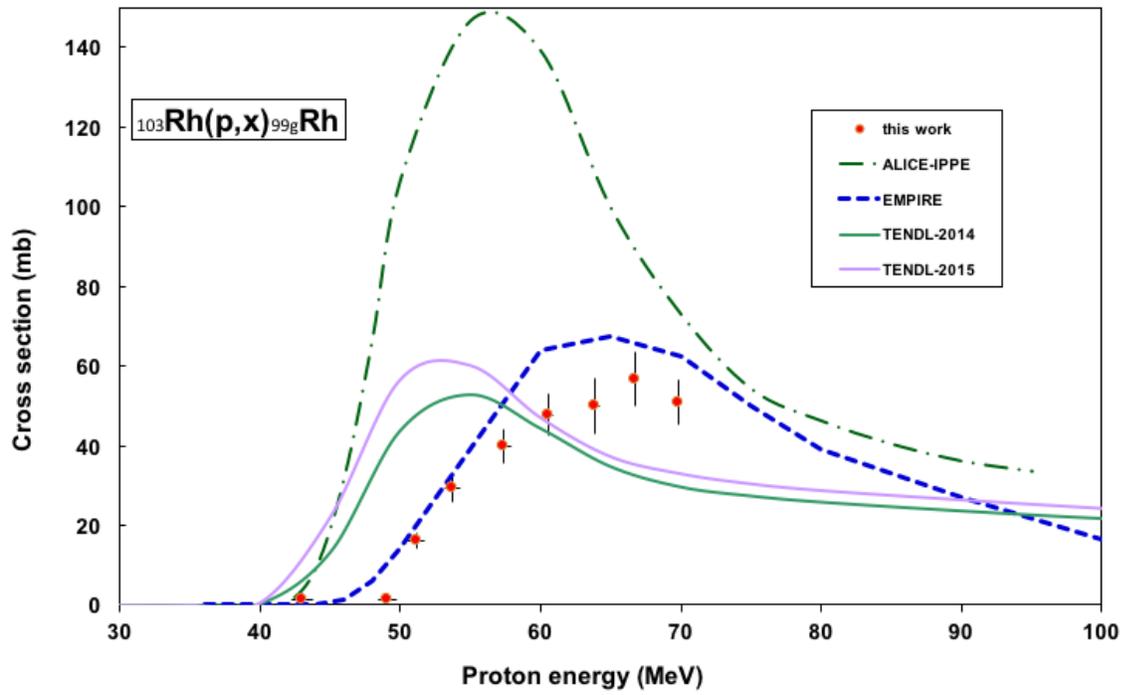

Fig.10. Excitation function of the $^{103}$Rh(p,x)$^{99g}$Rh reaction compared with the theory and literature

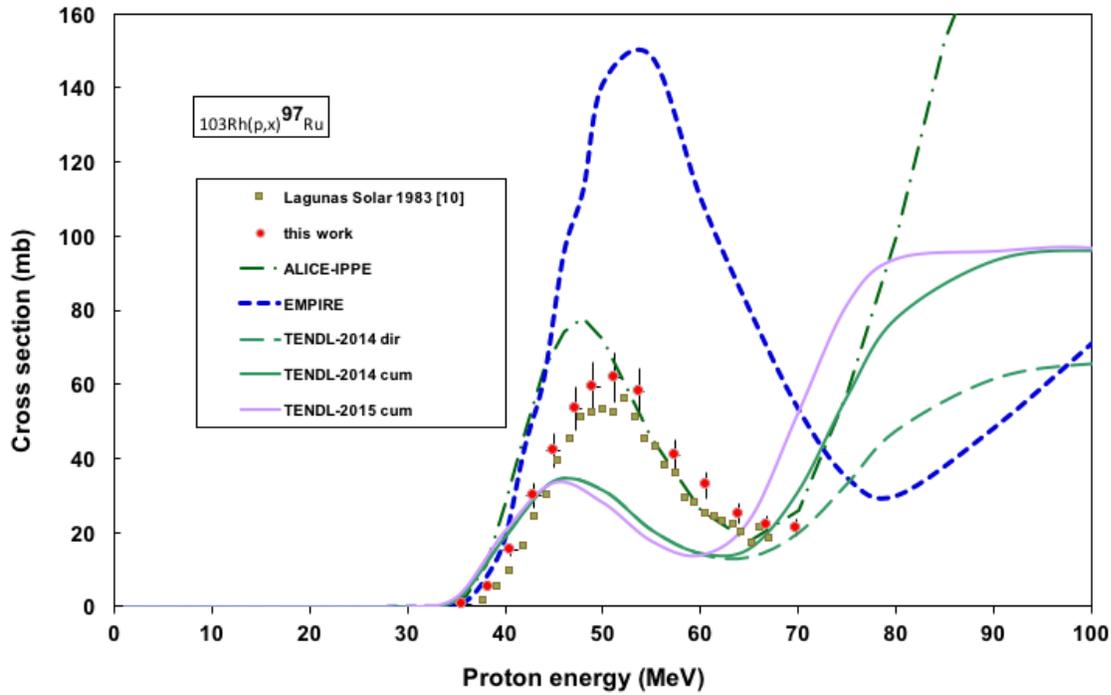

Fig.11. Excitation function of the $^{103}$Rh(p,x)$^{97}$Ru reaction compared with the theory and literature

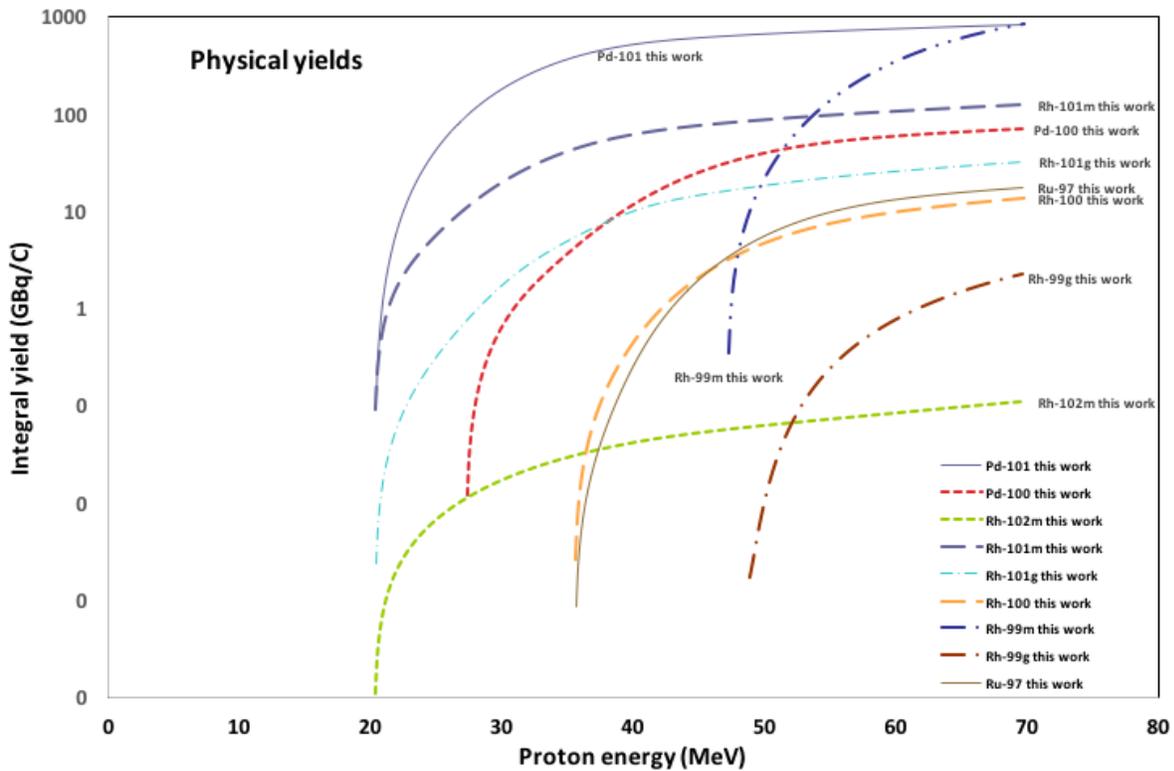

Fig. 12 Integral yields for the $^{101,100}$Pd, $^{102m, 102g,101m,101g,100,99m,99g}$Rh and $^{97}$Ru reactions